\documentclass[10pt,letterpaper]{article}
\usepackage{opex3}
\usepackage{amsfonts}
\usepackage{amsmath}
\usepackage{amssymb}
\usepackage{graphicx}
\usepackage{hyperref}

\begin{document}

\title{Rotating soliton solutions in nonlocal nonlinear media}

\author{S. Skupin and M. Grech}
\address{Max Planck Institute for the Physics of Complex Systems, N\"othnitzer Str.~38,
01187 Dresden, Germany}

\author{W. Kr\'olikowski}
\address{Laser Physics Center, Research School of Physical Sciences and Engineering,
Australian National University, Canberra, ACT 0200, Australia}

\begin{abstract}
We discuss generic properties of rotating nonlinear wave solutions,
the so called azimuthons, in nonlocal media. Variational methods allow us to derive
approximative values for the rotating frequency, which is shown to depend crucially
on the nonlocal response function. Further on, we link families of azimuthons
to internal modes of classical non-rotating stationary solutions, namely vortex and
multipole solitons. This offers an exhaustive method to identify azimuthons in a given nonlocal  medium.
\end{abstract}


\section{Introduction \label{intro}}

Spatial trapping of light in nonlinear media is a result of a balance between diffraction and the self-induced nonlinear
 index change  \cite{Stegeman:sc:286:1518,Kivshar:OS:03}. As the rate of diffraction is determined by the spatial scale
of the light beam independently of its dimensionality the stability properties of the self-trapped beams critically
depend on the particular model of the nonlinear response of the medium. Commonly encountered in the context of nonlinear
optics the, so called,  Kerr nonlinearity with nonlinear index change being  proportional to light intensity leads to
stable spatial solitons only in one transverse dimension. For two  dimensional beams the same model does not permit
stable beam localization. Instead, it predicts either eventual diffraction or catastrophic collapse  for input power
below or above a certain threshold value, respectively \cite{Berge:pr:303:259}. Hence, in order to ensure the existence
of stable self-localized beams in two dimensions different nonlinear response is necessary. It is known that  the
saturation of nonlinearity at high intensities is sufficient to prevent collapse and provide stabilization of finite
beams (see~\cite{Kivshar:pr:331:117} and references therein).
However, saturable nonlinearity cannot stabilize complex localized structures such as vortex beams which are prone to
azimuthal instability~\cite{Soto:pra:1991}. It has been shown recently that stabilization of localized waves is greatly
enhanced in the nonlocal nonlinear media. In such media nonlinear response in a particular spatial location is typically
determined  by the wave intensity in a certain neighborhood of this location. Nonlocality often results from certain
transport processes such as atomic diffusion~\cite{Suter:1993}, ballistic transport of hot atoms or
molecules~\cite{Skupin:2007:prl} or  heat transfer~\cite{Litvak:1975:sjpp,Rotschild:2006:np}. It can be also a signature
of a long-range interparticle interaction such as in nematic liquid
crystals~\cite{Assanto:2003:jqe,Conti:2004:prl,Peccianti:2006:prl}. Spatially nonlocal response is also naturally
present in  atomic condensates where it describes a  noncontact bosonic
interaction~\cite{Denschlag:2000,Pedri:2005:prl,Koch:2008:np}.
Extensive studies of beam propagation in nonlocal nonlinear media revealed a range of interesting features. In
particular, it has been shown that nonlocality may affect modulational instability of plane
waves~\cite{Krolikowski:pre:64:016612} and prevent catastrophic collapse of finite
beams~\cite{Turitsyn:1985,Bang:pre:66:046619} as well as stabilize complex one- and two-dimensional beams including
vortices~\cite{Briedis:2005:opex,Yakimenko:2005:pre,Skupin:2006,
Yakimenko:2006:pre,Kartashov:2006:ol,Kartashov:2007:oe}.  Recently, it has been shown that nonlocal media can also
support stable propagation of the rotating solitons, the so called azimuthons~\cite{Lopez-Aguayo:2006:ol}. Azimuthons
have been originally proposed  in the context of local nonlinear media~\cite{Desyatnikov:prl:95:203904}. They are 
azimuthally modulated beams  with nontrivial phase  structures exhibiting steady angular rotation upon propagation. They
can be considered as azimuthally perturbed optical vortices, i.e. beams with singular phase
structure~\cite{Coullet:1989}. Recent theoretical studies demonstrated both, stable and unstable evolution of
azimuthons~\cite{Buccoliero:2007:prl,Buccoliero:2008:ol}. In the former case it has been shown that in a highly nonlocal
regimes azimuthons undergo structural transformation resulting from the energetic coexistence of solitons of different
symmetries. Numerical and analytical studies revealed that the angular velocity of azimuthons is governed  by two
contributions. The linear component,  determined solely by the spatial structure of the beam (akin to rotation of
complex wave structures resulting from beating of their constituent modes~\cite{Abramochkin:2004:joapap}) as well as the
nonlinear which is brought about by the nonlinearity~\cite{Buccoliero:2008:ol}.

In this work we investigate in details  rotating nonlocal solitons. We consider various nonlocal models and show how
they determine dynamical properties of azimuthons. In particular, we will show that both, rotational frequency and
intensity profile of the azimuthons can be uniquely determined by analyzing Eigenmodes of
the linearized version of the corresponding nonlocal problem.
The paper is organized as follows: In Sec.~\ref{rotsol} we introduce the model equations, the general ansatz for
rotating solitons (azimuthons) and an exact expression for their rotation frequency. In Sec.~\ref{approtfreq}, we derive
an approximative, variational formula for the rotation frequency of the dipole azimuthons.
In Sec.~\ref{exrotfreq} we confront our semi-analytical results with exact numerical solutions of the ensuing
nonlinear equations. One of our main findings is that azimuthons emerge from internal modes of stationary nonlinear
soliton solutions like vortex or dipole, which is detailed in Sec.~\ref{intmodaz}.

\section{Rotating Solutions \label{rotsol}}

We consider physical systems governed by the two-dimensional nonlocal nonlinear Schr{\"o}dinger equation
\begin{equation}
\label{NLS}
i \frac{\partial}{\partial z} \psi +  \Delta_{\perp}\psi  + \theta \psi = 0.
\end{equation}
where $\theta$ represents the spatially nonlocal nonlinear response of the medium. Its form depends on the details of a
particular physical system.

In the so called Gaussian model of nonlocality, $\theta$ is given as
\begin{equation}
\label{NL_gaussian_int}
\theta = \frac{1}{2\pi}\iint\mathrm{e}^{-\frac{|\vec{r}-\vec{r}^{\prime}|^2}{2}}
\left|\psi(\vec{r}^{\prime},z)\right|^2d^2\vec{r}^{\prime},
\end{equation}
where $\vec{r}=x\vec{e}_x+y\vec{e}_y$ denotes the transverse coordinates.

If $\theta$ is governed by the following diffusion-type equation
\begin{equation}
\label{NL_thermal_diff}
\theta - \Delta_{\perp}\theta = |\psi|^2,
\end{equation}
solving formally Eq.\ (\ref{NL_thermal_diff}) in the Fourier space yields
\begin{equation}
\label{NL_thermal_int}
\theta = \frac{1}{2\pi}\iint\mathfrak{K}_0(|\vec{r}-\vec{r}^{\prime}|)
\left|\psi(\vec{r}^{\prime},z)\right|^2d^2\vec{r}^{\prime},
\end{equation}
where $\mathfrak{K}_0$ is the modified Bessel function of the second kind.

In the following, we will assume that the nonlinear response $\theta$ can be
expressed in terms of the nonlocal response function $R(r)$
\begin{equation}
\label{NL_R_int}
\theta = \iint R(|\vec{r}-\vec{r}^{\prime}|) \left|\psi(\vec{r}^{\prime},z)\right|^2d^2\vec{r}^{\prime},
\end{equation}
and Eqs.~(\ref{NL_gaussian_int}) and (\ref{NL_thermal_int}) will serve as
prominent examples.

Azimuthons are a straightforward generalization of the usual ansatz for
stationary solutions (solitons) \cite{Desyatnikov:prl:95:203904}.
They represent spatially rotating structures and hence involve an
additional parameter, the angular frequency $\Omega$
\begin{equation}
\label{azimuthon_ansatz}
\psi (r,\phi ,z) = U(r,\phi -\Omega z)\mathrm{e}^{i\lambda z},
\end{equation}
where $U$ is the complex amplitude function and $\lambda$ the propagation constant.
For $\Omega=0$, azimuthons become ordinary (nonrotating) solitons.
The most simple family of azimuthons is the one connecting the dipole soliton
with the single charged vortex soliton \cite{Lopez-Aguayo:2006:ol}. A single charged vortex
consists of two equal-amplitude
dipole-shaped structures representing real and imaginary part of $U$.
If these two components differ in amplitudes the resulting structure
 forms a ''rotating dipole'' azimuthon.  If one of
the components  is zero we deal with the (nonrotating) dipole soliton.
In the following we will denote the  ratio of these two amplitudes  by $\alpha$, which
also determines the angular modulation depth  of the resulting ring-like structure
by ``$1-\alpha$'' .

After inserting the  ansatz~(\ref{azimuthon_ansatz}) into Eqs.~(\ref{NLS}) and
(\ref{NL_R_int}), multiplying  with $U^*$ and $\partial_{\phi}U^*$ resp., and integrating
over the transverse coordinates we end up with
\begin{subequations}
\label{int_system}
\begin{align}
-\lambda M + \Omega L_z +I+N & =0 \label{int_system_a}\\
- \lambda L_z + \Omega M^{\prime} + I^{\prime} + N^{\prime} & =0.
\end{align}
This system relates the propagation constant $\lambda$ and the rotation
frequency $\Omega$ of the azimuthons to integrals over their stationary
amplitude profiles, namely
\begin{align}
M & = \iint \left|U(\vec{r})\right|^2d^2\vec{r} \\
L_z & = -i \iint U^*(\vec{r})
\frac{\partial}{\partial\phi}U(\vec{r})d^2\vec{r} \\
I & = \iint U^*(\vec{r})
\Delta_{\perp} U(\vec{r})d^2\vec{r} \\
N & = \iiiint R(|\vec{r}-\vec{r}^{\prime}|)
\left|U(\vec{r}^{\prime})\right|^2\left|U(\vec{r})\right|^2d^2\vec{r}^{\prime}d^2\vec{r}
\\
M^{\prime} & = \iint
\left|\frac{\partial}{\partial\phi}U(\vec{r})\right|^2d^2\vec{r} \\
I^{\prime} & = i \iint \left[ \frac{\partial}{\partial\phi} U^*(\vec{r}) \right]
\Delta_{\perp} U(\vec{r})d^2\vec{r} \\
N^{\prime} & = i \iiiint R(|\vec{r}-\vec{r}^{\prime}|)
\left|U(\vec{r}^{\prime})\right|^2 \left[ \frac{\partial}{\partial\phi}
U^*(\vec{r}) \right] U(\vec{r})d^2\vec{r}^{\prime}d^2\vec{r}.
\end{align}
\end{subequations}
The first two quantities have straightforward physical meanings, namely ''mass'' ($M$) and ''angular
momentum'' ($L_z$). We can formally solve for the rotation frequency and obtain (for an alternative
derivation see \cite{Rozanov:os:04})
\begin{equation}
\label{omega_exact}
\Omega=\frac{M\left(I^{\prime}+N^{\prime}\right)-L_z\left(I+N\right)}{L_z^2-MM^{\prime}}.
\end{equation}
Note that this expression is undetermined for a vortex beam (see discussion below).

While the above formula looks simple its use requires
detailed knowledge of the actual solution which can only be  obtained numerically.
However, it turns out that it is still possible to analyze main features of rotating solitons using
simple approximate approach.
To this end let us consider the ''rotating dipole''.
Obviously, we have $\Omega=0$ for $\alpha=0$ (dipole soliton). On the other
hand, for
$\alpha=1$ [vortex soliton $V(r)\exp(i\phi+i\lambda_0 z$)], we can assume any value for $\Omega$ by
just shifting the propagation constant $\lambda=\lambda_0+\Omega$ accordingly
($\lambda_0$ accounts for the propagation
constant in the non-rotating laboratory
frame). However, with
respect to the azimuthon in the limit $\alpha \rightarrow 1$, the value of $\Omega$ is
fixed.
In what follows, we denote this value by $\Omega|_{\alpha=1}$.

\section{Rotation Frequency - Approximate Analysis \label{approtfreq}}

In a first attempt, let us assume we know the radial shape of the vortex soliton
$V(r)$.
Then, a straight forward approximative ansatz for the dipole azimuthons is~\cite{Buccoliero:2007:prl,Buccoliero:2008:ol}
\begin{equation}
\label{azimuthon_trial}
U (r,\phi-\Omega z) = A V(r) \left[\cos (\phi-\Omega z)+i\alpha \sin (\phi-\Omega z) \right],
\end{equation}
where $A$ is an amplitude factor. The family of the ''rotating dipole''
azimuthons has two parameters, e.g. $\lambda$ and $\Omega$. Here, for technical
reasons, we prefer to use the mass $M$
and the amplitude ratio $\alpha$. We take the radial shape of the
vortex for a given mass $M_0$ and fix the amplitude factor $A$ in
ansatz~(\ref{azimuthon_trial}) to $A=\sqrt{2/(\alpha^2+1)}$. With this choice the mass of the azimuthon equals
$M_0$ for all values of $\alpha$. Then,
we can compute all integrals in Eqs.~(\ref{int_system}) as functions of
$\alpha$: $M=M_0$, $L_z=2\alpha M_0/(\alpha^2+1)$, $I=I_0$, $N=[4(1+\alpha ^4
) N_{\mathrm{cc}} + 8\alpha^2
N_{\mathrm{cs}}]/(\alpha^2+1)^2$, $M^{\prime}=M_0$, $I^{\prime}=2\alpha
I_0/(\alpha^2+1)$, and $N^{\prime}=4\alpha(N_{\mathrm{cc}}+N_{\mathrm{cs}})/(\alpha^2+1)$.
The index "$0$" indicates the value of the respective integral for the vortex (i.e., for $\alpha=1$),
and
\begin{subequations}
\begin{align}
N_{\mathrm{cc}} & = \iiiint R(|\vec{r}-\vec{r}^{\prime}|)
V^2(r^{\prime})\cos^2(\phi^{\prime}) V^2(r)\cos^2(\phi)d^2\vec{r}^{\prime}d^2\vec{r} \\
N_{\mathrm{cs}} & = \iiiint R(|\vec{r}-\vec{r}^{\prime}|)
V^2(r^{\prime})\cos^2(\phi^{\prime})  V^2(r)\sin^2(\phi)d^2\vec{r}^{\prime}d^2\vec{r}.
\end{align}
\end{subequations}
Finally, we obtain the following simple expression for the rotation frequency from
Eq.~(\ref{omega_exact}),
\begin{equation}
\label{omega_approx}
\Omega=\frac{2\alpha}{\alpha^2+1}\frac{2\left(N_{\mathrm{cc}}-N_{\mathrm{cs}}\right)}{M_0}.
\end{equation}

Let us have a closer look at this approximation. First of all, we have
$N_{\mathrm{cc}} \ge N_{\mathrm{cs}}$ if we assume $R(r)$ to be
monotonically decreasing in $r>0$.
This immediately implies that the sense of the amplitude rotation is opposite to that of the phase rotation.
In general, we need to know the
actual shape of the vortex soliton to compute Eq.~(\ref{omega_approx}).
One possibility would be to use exact numerical
solutions of the vortex shape $V$, a time consuming task because $V$ depends
in nontrivial way on $\lambda_0$. Moreover, Eq.~(\ref{omega_approx}) is only an
approximation to $\Omega$ and should not require the exact shapes. Luckily, it
is possible to compute quite good approximative vortex solitons using the
so called variational approach \cite{anderson}. We employ the following ansatz $V=Ar\exp(-r^2/2\sigma^2)$,
$U=V\exp(i\phi)$, $\psi=U\exp(i\lambda_0z)$ and look for critical points of the Lagrangian
$\mathcal{L}=-\lambda_0M+I+N/2$ with respect to the variables $A$ and
$\sigma$. In the following, we give results for our two exemplary nonlocal
nonlinearities.

\begin{figure}
\centering\includegraphics[width=8cm]{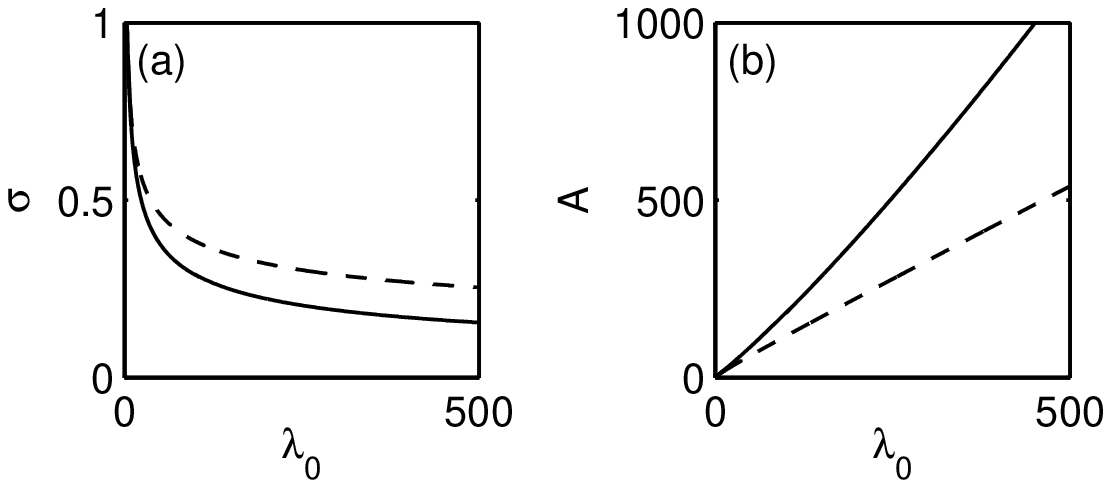}
\centering\includegraphics[width=8cm]{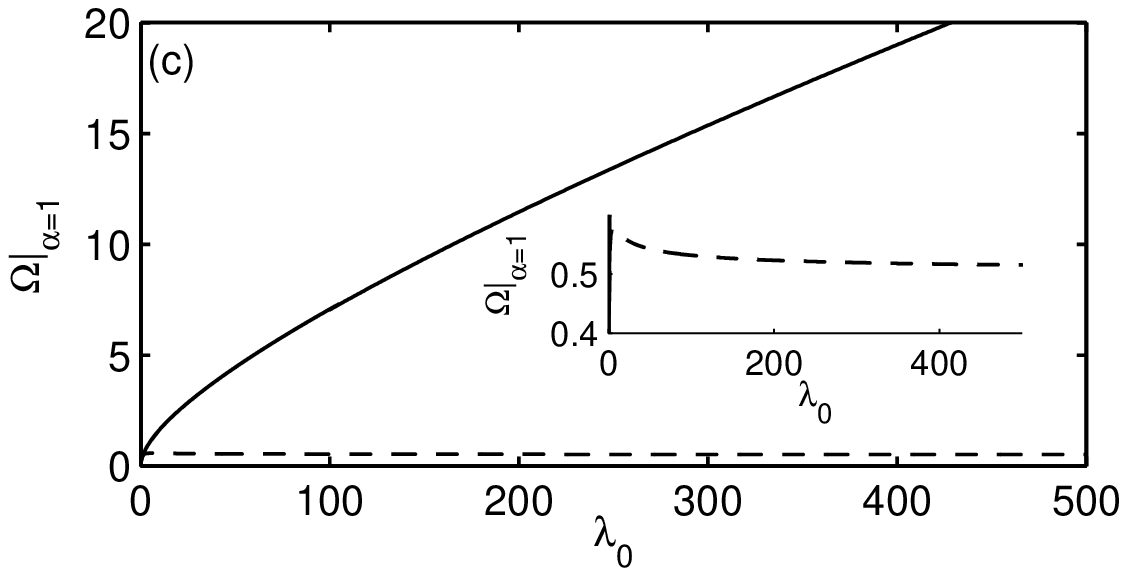}
\caption{\label{fig_var} Nonlocal model and variational
approach: Solid lines show (a) width $\sigma$, (b) amplitude $A$, and (c)
frequency
$\Omega|_{\alpha=1}$ for the Bessel response as functions of the propagation
constant $\lambda_0$. Dashed
lines represent the same quantities for the Gaussian nonlocal response.
The inset in (c) shows $\Omega|_{\alpha=1}$ for the latter response on more appropriate scales.
 While both,  width $\sigma$ and amplitude $A$,
exhibit similar behavior  for the two nonlocal models
 the rotation frequency $\Omega|_{\alpha=1}$
shows a completely different dependence on $\lambda_0$ and drastic
difference in  magnitude for Gaussian and Bessel responses.}
\end{figure}

For the Gaussian response, Eq.~(\ref{NL_gaussian_int}), all integrals appearing
in the variational approach can be evaluated analytically. Hence, we find the
width $\sigma$ of our approximate solution as
\begin{equation}
\label{sigma_gauss}
\lambda_0 \sigma^8+\left(-2+2\lambda_0 \right) \sigma^6+\left(-8+4\lambda_0 \right)\sigma^4-8\sigma^2-8=0,
\end{equation}
its amplitude
\begin{equation}
\label{A_gauss}
A=\frac{2\left(\sigma^2+1\right)}{\sigma^3}\sqrt{\frac{\left(\sigma^2+1\right)\left(\lambda_0\sigma^2+2\right)}{
\sigma^4+2\sigma^2+2}},
\end{equation}
and the rotation frequency $\Omega$
\begin{equation}
\label{omega_gauss}
\Omega=\frac{2\alpha}{\alpha^2+1}\frac{A^2\sigma^8}{8\left(\sigma^2+1\right)^3}.
\end{equation}
We show these three quantities as functions of $\lambda_0$ in
Fig.~\ref{fig_var} (dashed lines). If $\lambda_0$ is large, $\sigma\ll1$
(nonlocal limit) and we can
neglect the first two terms in  Eq.~(\ref{omega_gauss}). Then we get $\sigma_{\rm
nl}=(2/\lambda_0)^{1/4}$, $A_{\rm nl}=\lambda_0(\sigma_{\rm nl}^2+1)^{3/2}$
and, subsequently, $\Omega_{\rm nl}=\alpha/(\alpha^2+1)$. The immediate  conclusion drawn
from this formula  is that dipole azimuthons in the
Gaussian nonlocal model do not rotate  faster than $\Omega\sim0.5$. Moreover,
in the nonlocal regime the rotating frequency does not change much with the propagation
constant $\lambda$ or the mass $M$ of the azimuthon, but is mainly determined by
 the structural parameter $\alpha$.

For the Bessel nonlocal response, Eq.~(\ref{NL_thermal_int}), we have to compute the
relevant integrals 
numerically. The results for both, Bessel and Gaussian  models are  presented in Fig.~\ref{fig_var} using 
solid and  dashed  lines, respectively.  It is clear that amplitude $A$ as well as soliton width $\sigma$, follow
similar dependencies as a function of the propagation constant  $\lambda_0$ [see Fig.~\ref{fig_var}(a-b)].
 Totally  different dependence shows the rotation frequency $\Omega$ [see Fig.~\ref{fig_var}(c)]. In particular, not
only azimuthons
in the Bessel nonlocal model rotate much faster
than those predicted by the Gaussian model but also the former exhibit strong sensitivity of $\Omega$ on 
 the propagation constant $\lambda$ and mass $M$.

\begin{figure}
\centering\includegraphics[width=8cm]{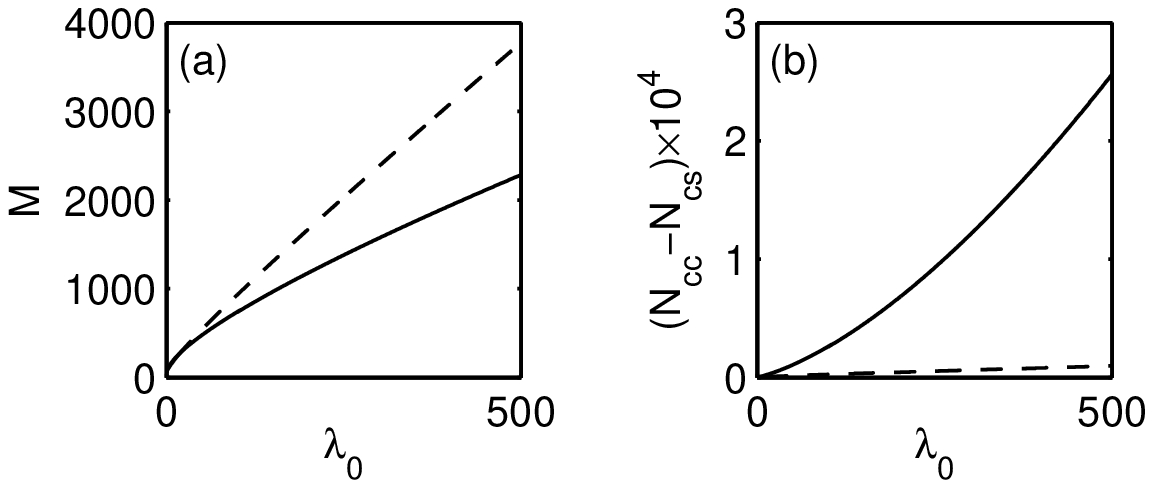}
\includegraphics[width=4cm]{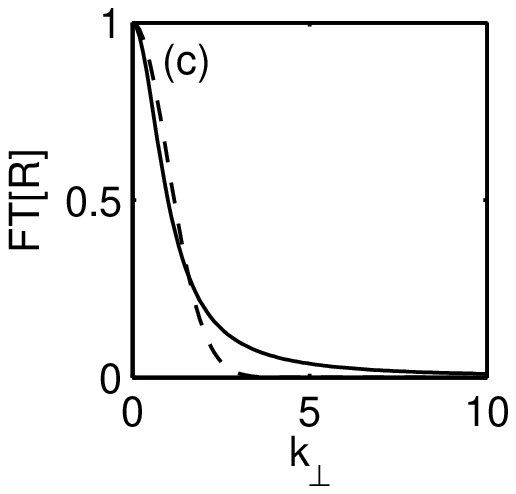}
\caption{\label{fig_omega_ingr} (a) Mass $M$ and (b) quantity
$N_{\mathrm{cc}}-N_{\mathrm{cs}}$ for Bessel (solid lines) and Gaussian (dashed
lines) nonlocal responses as functions of the propagation
constant $\lambda_0$. It is obvious that the difference in the rotating frequency $\Omega|_{\alpha=1}$
shown in Fig.~\ref{fig_var}(c) is due to $N_{\mathrm{cc}}-N_{\mathrm{cs}}$ [see Eq.~(\ref{omega_approx})].
(c) Spatial Fourier Transform (FT) of the Bessel (solid lines) and Gaussian (dashed
lines) nonlocal response $R$ as a function of the transverse wave-vector $k_{\perp}=\sqrt{k_x^2+k_y^2}$.
The degree of
spatial averaging is obviously larger for the Gaussian than for the Lorentzian.}
\end{figure}

To shed light on the origin of the difference in the  rotating frequencies of  the two
models let us have a look at the quantities $M$ and
$N_{\mathrm{cc}}-N_{\mathrm{cs}}$, which directly determine
$\Omega|_{\alpha=1}$ in the present approximation [see
Eq.~(\ref{omega_approx})].
Figures~\ref{fig_omega_ingr}(a,b) reveal that as the masses $M$ are comparable for
both  models, the important differences come from $N_{\mathrm{cc}}-N_{\mathrm{cs}}$. Hence, we
can conclude that it is not the actual shape of the azimuthon which determines
$\Omega$, but the convolution integrals $N_{\mathrm{cc}}$ and
$N_{\mathrm{cs}}$ with the response function. In fact, $N_{\mathrm{cc}}$ and
$N_{\mathrm{cs}}$ represent the overlaps of the nonlocal response of one dipole
component with either itself or its $\pi/2$-rotated counterpart. Hence, the
more of the non-rotational-symmetric shape is preserved by the nonlocal
response, the larger is $N_{\mathrm{cc}}-N_{\mathrm{cs}}$. 
Looking at both response functions considered here in Fourier domain [Fig.~\ref{fig_omega_ingr}(c)] it is obvious that
the degree of nonlocality and consequently the degree of spatial averaging is greater for the Gaussian response leading
to smaller difference between $N_{\mathrm{cc}}$ and $N_{\mathrm{cs}}$.

\section{Rotation Frequency - Numerical Analysis\label{exrotfreq}}

\begin{figure}
\centering\includegraphics[width=8cm]{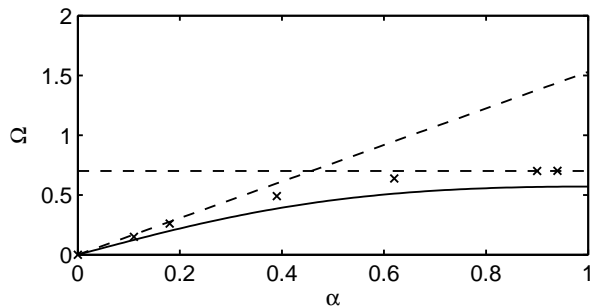}
\caption{\label{fig_azim_gauss_200} Gaussian nonlocal model, $M=200$,
$\lambda_0=11.5$: Solid line shows the result of Eq.~(\ref{omega_approx})
computed from the variational vortex approximation.
Crosses depict results obtained from direct numerical solutions. Dashed
lines represent results of the  linear perturbation analysis of dipole and vortex
soliton, respectively (see Sec.~\ref{intmodaz}).}
\end{figure}

\begin{figure}
\centering\includegraphics[width=8cm]{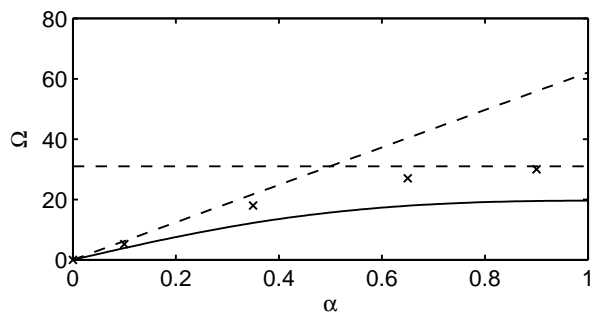}
\caption{\label{fig_azim_bessel_2000} Bessel nonlocal model, $M=2000$,
$\lambda_0=418$: Same coding as in Fig.~\ref{fig_azim_gauss_200}.}
\end{figure}

How good are our approximations of the rotating frequency $\Omega$? Is the
dependency in the structural parameter $\alpha$ really approximatively
$\alpha/(\alpha^2+1)$ for a given $\lambda_0$?
To answer these questions we compute the azimuthons $U$ numerically
and propagate them over some distance in $z$ to observe their rotation and
estimate $\Omega$. One typical example for the Gaussian nonlocal model is shown in
Fig.~\ref{fig_azim_gauss_200}. The corresponding vortex soliton has a
propagation constant $\lambda_0=11.5$ and mass $M=200$. As its  width $\sigma\approx 0.7$ 
is smaller than the width of the response
function   we are definitely in the nonlocal regime. We can clearly see the reasonable
agreement between approximate   and fully numerical   results [compare with Fig.~\ref{fig_var}(c)].
We also tested azimuthons with larger masses and therefore corresponding to higher degrees of nonlocality.
For instance, for $\lambda_0=247$, $M=2000$ and $\alpha=0.95$ we find $\Omega=0.54$ from direct numerical simulations,
in excellent agreement with $\Omega=0.52$ from the approximate model [see also
Fig.~\ref{fig_var}(c)].
Figure~\ref{fig_azim_bessel_2000} shows another example, this time  for the Bessel
nonlocal model. Again, the agreement is reasonable. Here, we have to use higher
$\lambda_0$ and larger mass in order to ensure the stability of the  vortex soliton. Like
in the previous example, we operate in a nonlocal regime ($\sigma=0.17$).
To summarize,  our simple approximate 
model for $\Omega$, which involves nothing but a variational solution for the
vortex soliton, allows us to predict the correct range and sign of $\Omega$ for a
family of dipole azimuthons with given mass. Moreover, the dependency on the
structural parameter $\alpha$ is also correctly modeled.

However, there is still serious  discrepancy in the values of $\Omega|_{\alpha=1}$. One could
think of two possible reasons. Firstly, it could be  the deviations between the actual vortex soliton
and its variational approximation.  However, it turns out that   
$2(N_{\mathrm{cc}}-N_{\mathrm{cs}})/M$ computed from the numerically obtained vortex
profile  is   almost constant (within a few percents). Hence it has no effect on  the overall error in
$\Omega|_{\alpha=1}$.  
However, postulating the specific  ansatz Eq.~(\ref{azimuthon_ansatz})
we implicitly assumed a certain shape of the deformation to the soliton profile while
going over  from the vortex to azimuthons, in the limit  $\alpha \rightarrow 1$. Therefore it has
to be this shape of vortex deformation which determines the rotation frequency $\Omega$, since a
vortex formally allows for all possible rotations (see discussion on shifting
$\lambda$ in end of Sec.~\ref{rotsol}). In the next section we will therefore discuss  the formation 
of azimuthons by considering it as a  process of bifurcation from the stationary non-rotating 
soliton solutions.

\section{Internal Modes and Azimuthons \label{intmodaz}}

Let us start with the dipole soliton $D(r,\phi)$, because here the resulting linear
problem is simpler than in the vortex case. We can directly use the rotation
frequency $\Omega$ as the linearization parameter, since for the dipole we have
$\Omega_{\mathrm{D}}=0$. We use  the following ansatz
\begin{equation}
\label{dipole_ansatz}
U (r,\phi-\Omega z) = D(r,\phi-\Omega z) + i \Omega d(r,\phi-\Omega z),
\end{equation}
where $D$ and $d$ are real functions representing soliton and its small deformation, respectively.
 We then  plug this ansatz into Eqs.~(\ref{NLS}), (\ref{NL_R_int})
and (\ref{azimuthon_ansatz}) and linearize with respect to $\Omega$. The resulting equation of order
$\mathcal{O}(\Omega^0)$ is fulfilled by the dipole soliton $D$. The terms of the order
$\mathcal{O}(\Omega^1)$ lead to the following  linear equation for the deformation $d$:
\begin{equation}
\label{dipole_equation}
-\lambda_{\mathrm{D}} d + \Delta_{\perp}d + \iint R(|\vec{r}-\vec{r}^{\prime}|)
D^2(\vec{r}^{\prime})d^2\vec{r}^{\prime} d =  \frac{\partial}{\partial\phi}D.
\end{equation}
In deriving the above equation we made use of the fact that  the propagation constant $\lambda$ of the
azimuthon depends on $\Omega$ as $\lambda=\lambda_{\mathrm{D}} +
\mathcal{O}(\Omega^2)$, where $\lambda_{\mathrm{D}}$ is the propagation
constant of the dipole soliton $D$. This can be seen easily from 
Eq.~(\ref{int_system_a}) and $M=M_{\mathrm{D}}+\mathcal{O}(\Omega^2)$,
$L_z=0+\mathcal{O}(\Omega)$, $I=I_{\mathrm{D}}+\mathcal{O}(\Omega^2)$, and
$N=N_{\mathrm{D}}+\mathcal{O}(\Omega^2)$, where the index ''${\mathrm{D}}$''
indicates the respective integral for the dipole soliton $D$. 

\begin{figure}
\centering\includegraphics[width=12cm]{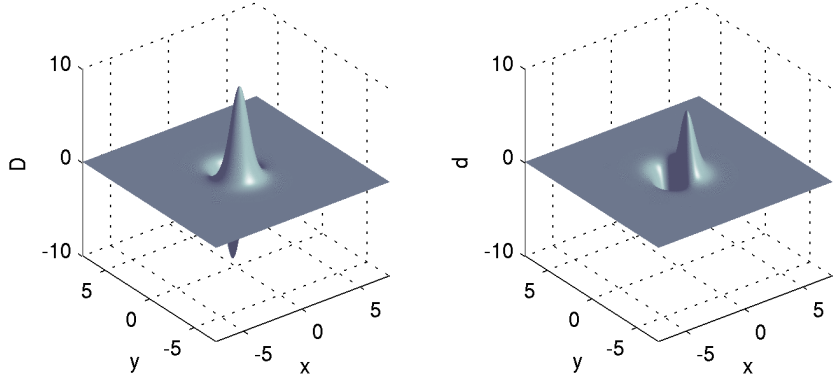}
\caption{\label{fig_diplole_gauss_200} Dipole $D$ and deformation $d$ for the
Gaussian nonlocal model, $M=200$, $\lambda_D=11.8$. The deformation $d$ has a dipole shape, rotated by $\pi/2$ with
respect to $D$.
We can construct an approximate azimuthon $U = D+i \Omega d$ for small rotation frequencies $\Omega$. The amplitude
ratio $\max d / \max D$
gives the slope of $\Omega(\alpha)$ at $\alpha=0$.}
\end{figure}

The solution of Eq.(\ref{dipole_equation}) for the
Gaussian nonlocal model and the dipole soliton with $M=200$,
and $\lambda_D=11.8$ is shown in Figure~\ref{fig_diplole_gauss_200}.
The left plot shows  the dipole soliton $D(x,y)$ while the right shows its deformation $d(x,y)$.
As we can see from the right subplot,
the deformation $d$ has a  shape of a dipole as well,
but rotated by $\pi/2$ with respect to $D$. The width of $d$ is about $0.85$
that of $D$. Hence,  for small rotation frequency $\Omega$ the dipole azimuthon
consists of two orthogonally oriented dipoles,  
with the the relative phase shift of $\pi/2$ and amplitude ratio
$\alpha = \Omega \max d / \max D$. In effect, the whole rotating structure is slightly elliptic, 
with the principal axis given by $D$. In our example, we find $\max d / \max D = 1.53$,
hence $\Omega = 1.53 \alpha$,
which agrees perfectly with exact numerical results obtained from the solution of the original nonlinear problem
Eq.(\ref{NLS})  for   $\alpha \rightarrow 0$ (see Fig.~\ref{fig_azim_gauss_200}).

\begin{figure}
\centering\includegraphics[width=12cm]{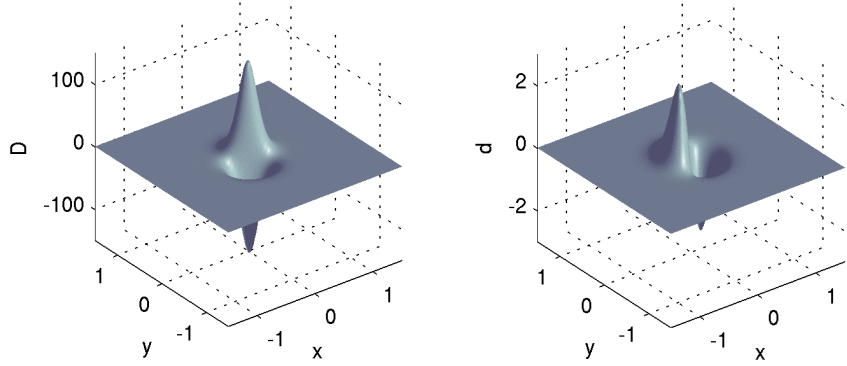}
\caption{\label{fig_dipole_bessel_2000}Same as Fig.~\ref{fig_diplole_gauss_200} but for the
Bessel nonlocal model, $M=2000$, $\lambda_D=444$.}
\end{figure}

We obtain similar results for our second example, the dipole soliton in the
Bessel nonlocal model with $M=2000$, $\lambda_D=444$ (see Fig.~\ref{fig_dipole_bessel_2000}). 
The ratio of the widths of a dipole $D$ and its deformation $d$
is $0.76$, and $\Omega = 62.1 \alpha$. Again, the agreement  with the
full numerical results of Fig.~\ref{fig_azim_bessel_2000} is excellent.

Let us now look at the  azimuthon originating (bifurcating) from the vortex soliton.   
 For this purpose, we recall the Eigenvalue problem
for the internal modes of the  nonlinear potential $\theta$ which is usually treated in the context of 
 linear stability of nonlinear (soliton) solutions. We introduce a small 
 perturbation $\delta V$ to the vortex soliton $V$,
and plug
\begin{equation}
\psi = \left(V+\delta V\right)\mathrm{e}^{i\phi+i\lambda_0z}
\end{equation}
into Eqs.~(\ref{NLS}) and
(\ref{NL_R_int}) and linearize with respect to the perturbation.
Please note that the perturbation $\delta V(r,\phi,z)$ is complex, whereas the vortex profile $V(r)$ can be chosen real.
The resulting evolution equation for the perturbation $\delta V$ is given by
\begin{equation}
\begin{split}
\left[ i \frac{\partial}{\partial z} -\lambda_0 +
\frac{1}{r}\frac{\partial}{\partial r}\left(r\frac{\partial}{\partial
r}\right) + \frac{1}{r^2} \left(\frac{\partial}{\partial \phi} +i
\right)^2 + \iint R(|\vec{r}-\vec{r}^{\prime}|) V^2(r^{\prime})d^2\vec{r}^{\prime} \right]\delta V & \\
+ V\iint R(|\vec{r}-\vec{r}^{\prime}|) V(r^{\prime}) \left[ \delta V(\vec{r}^{\prime},z) + \delta
V^*(\vec{r}^{\prime},z) \right] d^2\vec{r}^{\prime} &=0.
\end{split}
\end{equation}
With the ansatz
\begin{equation}
\delta V = \delta V_1(r)\mathrm{e}^{im\phi+i\kappa z} + \delta V^*_2(r)\mathrm{e}^{-im\phi-i\kappa^* z}
\end{equation}
we then derive the Eigenvalue problem for the internal modes
\begin{equation}
\label{Eigen_vortex}
\hat{L} \begin{pmatrix} \delta V_1\\ \delta V_2\end{pmatrix}
= \kappa \begin{pmatrix} \delta V_1\\ \delta V_2\end{pmatrix},
\end{equation}
where \footnote{Please not that since
$|\vec{r}-\vec{r}^{\prime}|=\sqrt{r^2+r^{\prime2}
-2rr^{\prime}\cos(\phi-\phi^{\prime})}$,
all integrals in (\ref{linop}) are independent of $\phi$.}
\begin{equation}
\label{linop}
\begin{split}
\hat{L} \begin{pmatrix} \delta V_1\\ \delta V_2\end{pmatrix}
& = \begin{pmatrix}
 \left[ \frac{1}{r}\frac{\partial}{\partial r}\left(r\frac{\partial}{\partial r}\right) - \frac{(m+1)^2}{r^2}
 -\lambda_0 + \iint R(|\vec{r}-\vec{r}^{\prime}|) V^2(r^{\prime})d^2\vec{r}^{\prime} \right] \delta V_1 \\
 -\left[\frac{1}{r}\frac{\partial}{\partial r}\left(r\frac{\partial}{\partial r}\right) - \frac{(m-1)^2}{r^2}
 -\lambda_0 + \iint R(|\vec{r}-\vec{r}^{\prime}|) V^2(r^{\prime})d^2\vec{r}^{\prime} \right] \delta V_2
\end{pmatrix} \\
& \quad +\begin{pmatrix}
V\iint R(|\vec{r}-\vec{r}^{\prime}|)V(r^{\prime}) \left[\delta V_1(r^{\prime}) + \delta V_2(r^{\prime}) \right]
\cos[m(\phi-\phi^{\prime})] d^2\vec{r}^{\prime} \\
-V\iint R(|\vec{r}-\vec{r}^{\prime}|)V(r^{\prime}) \left[\delta V_2(r^{\prime}) +\delta V_1(r^{\prime}) \right]
\cos[m(\phi-\phi^{\prime})] d^2\vec{r}^{\prime}
\end{pmatrix}.
\end{split}
\end{equation}
Real Eigenvalues of Eq.~(\ref{Eigen_vortex}) ($\kappa=\kappa^*$) are termed
orbitally stable and the corresponding Eigenvector $(\delta V_1, \delta V_2)$
can be chosen as real function.
If we perturb the vortex $V$ with an orbitally stable
Eigenvector, the resulting wave-function $\psi$ can be written in the form of
Eq.~(\ref{azimuthon_ansatz}) with $\Omega=-\kappa/m$ and $\lambda=\lambda_0-\kappa/m$.
Hence, we expect to find an orbitally stable internal mode with $m=2$ corresponding to
the azimuthon, where the Eigenvalue $\kappa$ gives $\Omega|_{\alpha=1}=-\kappa/2$.

\begin{figure}
\centering\includegraphics[width=6cm]{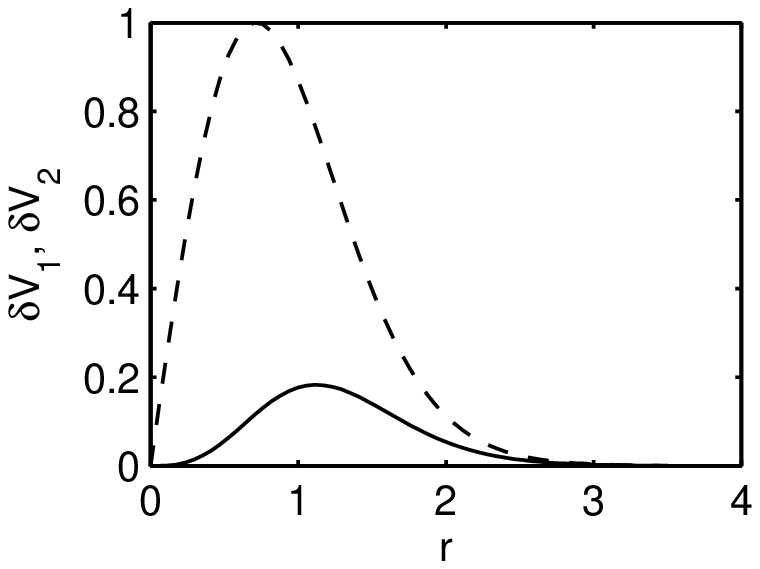}
\centering\href{http://www.pks.mpg.de/~skupin/azimuthon/movie1_short.avi}{\includegraphics[width=12cm]{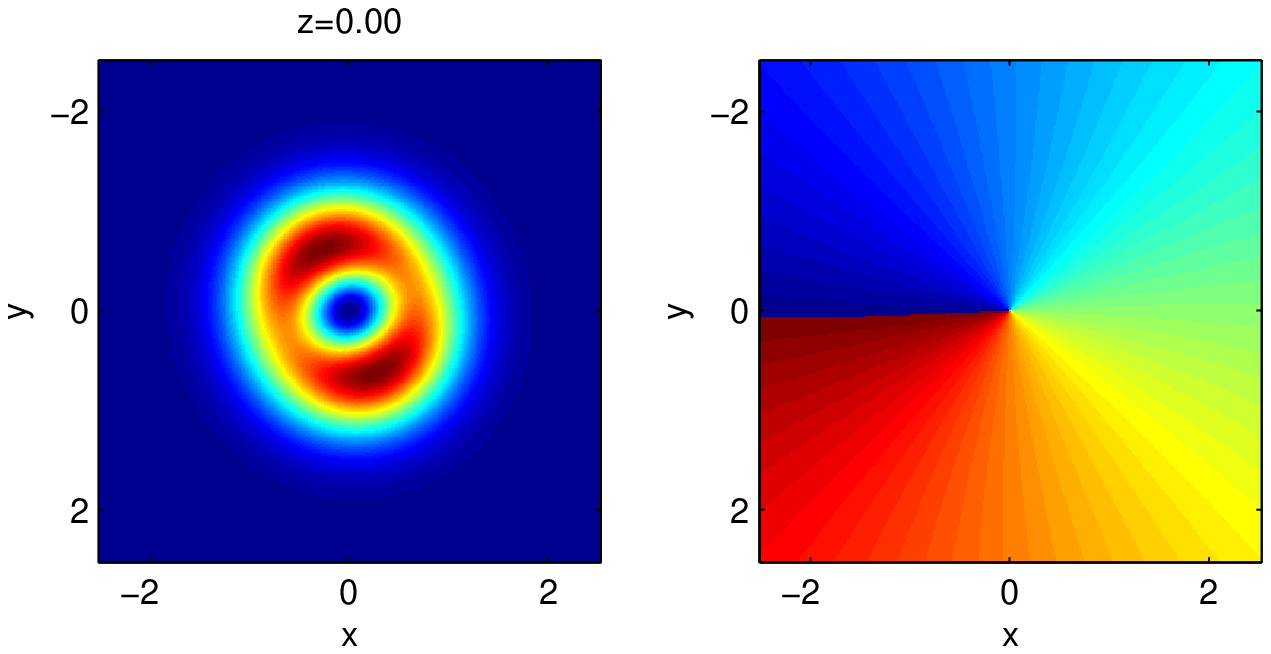}}
\caption{\label{fig_azimvor_gauss} Gaussian nonlocal model, $M=200$,
$\lambda_0=11.5$: Solid line shows the component $\delta V_1$, dashed line
the component $\delta V_2$, for $m=2$ and Eigenvalue $\kappa=-1.4$
computed from Eq.~(\ref{Eigen_vortex}).
The movie (\href{http://www.pks.mpg.de/~skupin/azimuthon/movie1_short.avi}{4~MB} or
\href{http://www.pks.mpg.de/~skupin/azimuthon/movie1_long.avi}{15~MB}) shows the evolution of intensity and
phase of the vortex perturbed with the Eigenvector shown
above,
$U|_{z=0}=V\exp(i\phi) + \delta V_1 \exp(i 3 \phi) + \delta V_2 \exp(-i \phi)$. The chosen amplitudes of the
perturbation are such that $\alpha=0.85$,
the rotating frequency is close to $\Omega|_{\alpha=1}=0.7$.}
\end{figure}

\begin{figure}
\centering\includegraphics[width=6cm]{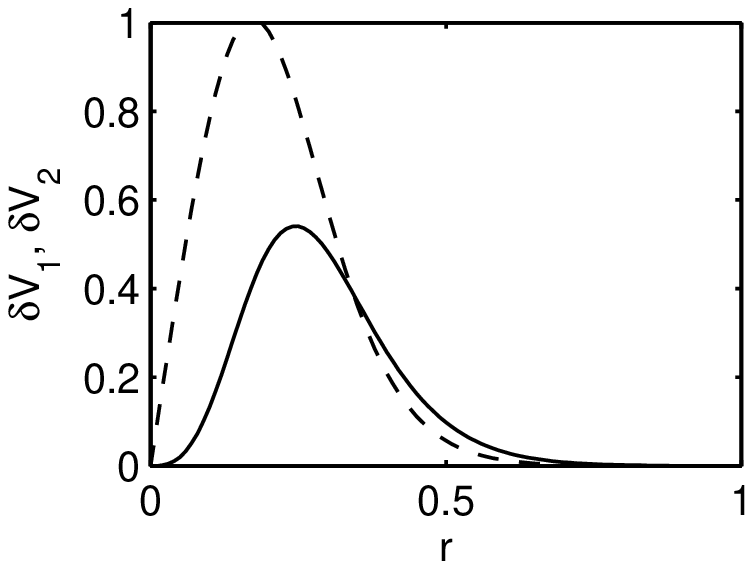}
\centering\href{http://www.pks.mpg.de/~skupin/azimuthon/movie2_short.avi}{\includegraphics[width=12cm]{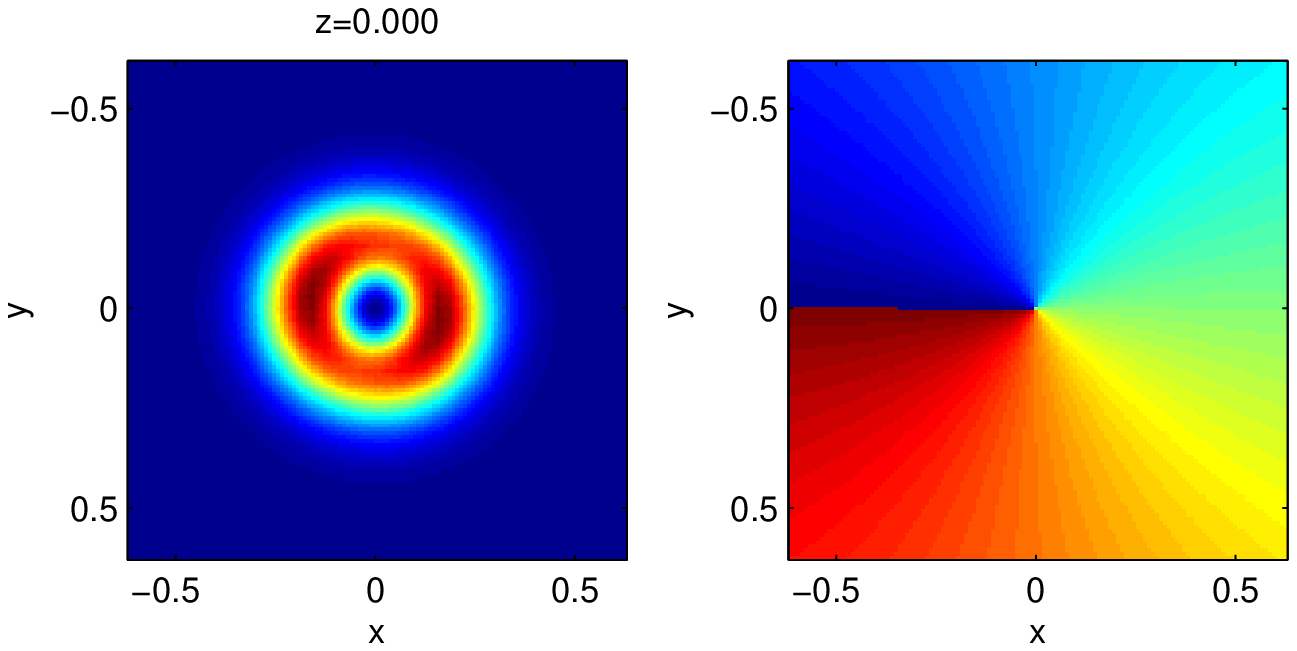}}
\caption{\label{fig_azimvor_bessel} Bessel nonlocal model, $M=2000$,
$\lambda_0=418$: Same as in Fig.~\ref{fig_azimvor_gauss}, $m=2$, $\kappa=-62$.
The movie (\href{http://www.pks.mpg.de/~skupin/azimuthon/movie2_short.avi}{4~MB} or
\href{http://www.pks.mpg.de/~skupin/azimuthon/movie2_long.avi}{15~MB}) shows the evolution of the perturbed
vortex with $\alpha=0.9$, the rotating frequency is close
to
$\Omega|_{\alpha=1}=31$.}
\end{figure}

Indeed, when we solve problem~(\ref{Eigen_vortex}) for our two test cases
numerically we find Eigenvalues at the expected positions, $\kappa=-1.4$ and
$\kappa=-62$ respectively. The resulting estimates for $\Omega|_{\alpha=1}$
fit perfectly with values obtained from direct simulations (see
Figs.~\ref{fig_azim_gauss_200} and \ref{fig_azim_bessel_2000}). The radial
profiles of the corresponding Eigenvectors are shown in
Figs.~\ref{fig_azimvor_gauss} and \ref{fig_azimvor_bessel}). The total
vorticity of the component $\sim \delta V_1$ is $2+1=3$, the one of $\sim
\delta V_2$ is $-2+1=-1$. These total vorticities manifest themselves in the
shapes of the radial profiles near the origin $r=0$.

As for the perturbation of the stationary dipole, it is possible to 
construct azimuthons in the vicinity of the vortex ($\alpha \approx 1$)
from $\delta V$:
\begin{equation}
\label{azimvor}
U (r,\phi)|_{z=0} = \left[ V(r) + \delta V_1(r)\mathrm{e}^{im\phi} + \delta
V_2(r)\mathrm{e}^{-im\phi}\right]\mathrm{e}^{i\phi}.
\end{equation}
Used as an initial condition  to the propagation equation this object is expected to rotate with
$\Omega|_{\alpha=1}=-\kappa/m$.
Unlike in the previous case of the stationary dipole, the amplitude of the perturbation is not fixed\footnote{Just the
ratio between the components $\delta V_1$ and $\delta V_1$ is fixed.}, but will eventually determine the value of the
structural parameter $\alpha$. Generally speaking, the smaller the resulting $\alpha$ the greater the error in the
constructed initial condition. However, the great robustness of the azimuthons  allows one to use the initial
condition~(\ref{azimvor}) for quite large perturbation amplitudes (resulting $\alpha\sim0.5$). Those ''bad'' initial
conditions result in  oscillations of the azimuthon upon propagation. However, the azimuthon is structurally stable and does not
decay into other solutions like the single-hump ground state. Moreover, such initial conditions play a role of 
excellent ''initial guesses'' for solver routines to find numerically exact azimuthons. In the movies shown in
Figs.~\ref{fig_azimvor_gauss} and \ref{fig_azimvor_bessel} we choose moderate amplitudes of the perturbations in order
to get almost perfect azimuthons. The corresponding structural parameters $\alpha\ge0.85$ also
guarantee that the observable rotating frequencies are very close to their limiting values $\Omega|_{\alpha=1}$.

The results of the linear stability analysis of the vortex soliton offer an easy
explanation for the observed systematic deviation of the estimation for
$\Omega|_{\alpha=1}$ by Eq.~(\ref{omega_approx}) from its true value. As
mentioned before, the shape of the perturbation to the vortex soliton determines
$\Omega|_{\alpha=1}$. In our ansatz (\ref{azimuthon_trial}) we only account
for a perturbation similar to the component $\sim \delta V_2$ and ''forget''
the other component  $\sim \delta V_1$ with total vorticity 3, which leads to the observed
deviation. The larger the ''forgotten'' component, the larger is this deviation
(compare Figs.~\ref{fig_azimvor_gauss}, \ref{fig_azim_gauss_200} and
Figs.~\ref{fig_azimvor_bessel}, \ref{fig_azim_bessel_2000}).

\begin{figure}
\centering\includegraphics[width=6cm]{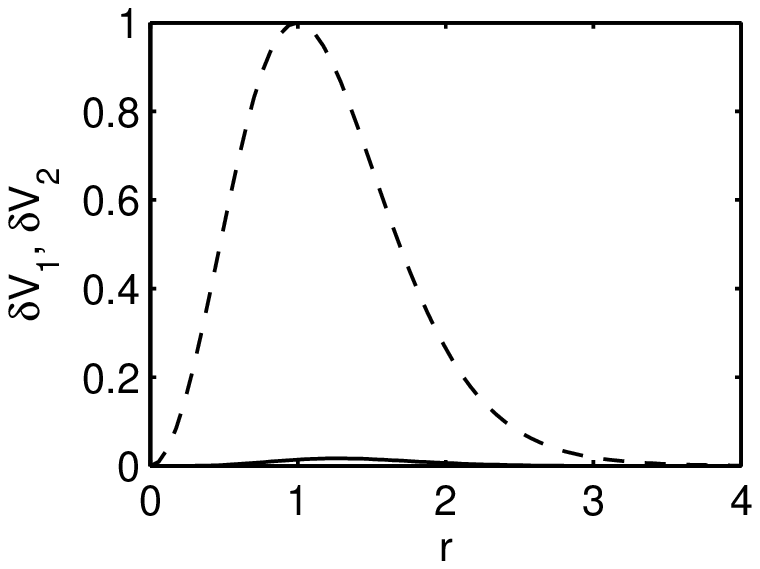}
\centering\href{http://www.pks.mpg.de/~skupin/azimuthon/movie3_short.avi}{\includegraphics[width=12cm]{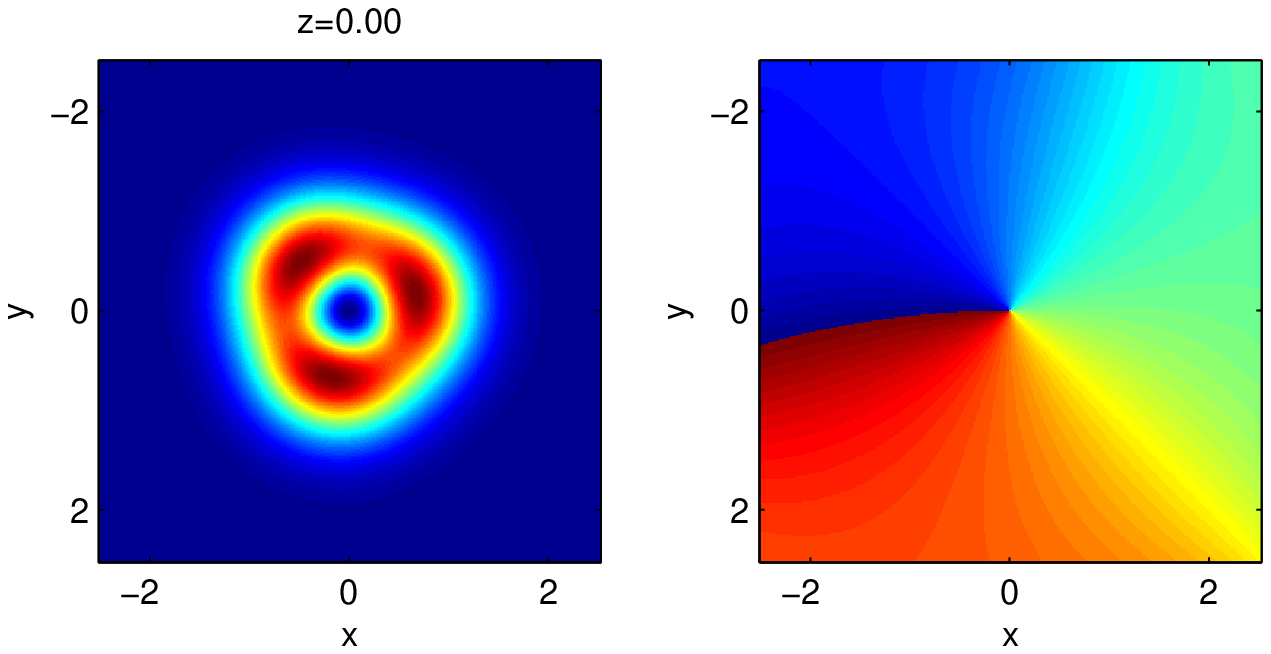}}
\caption{\label{fig_azim3vor_gauss}Gaussian nonlocal model, $M=200$,
$\lambda_0=11.5$: Same as in Fig.~\ref{fig_azimvor_gauss}, $m=3$,
$\kappa=3.8$.
The movie (\href{http://www.pks.mpg.de/~skupin/azimuthon/movie3_short.avi}{4~MB} or
\href{http://www.pks.mpg.de/~skupin/azimuthon/movie3_long.avi}{15~MB}) shows the evolution of perturbed
vortex
$U|_{z=0}=V\exp(i\phi) + \delta V_1 \exp(i 4 \phi) + \delta V_2 \exp(-i 2 \phi)$, $\alpha=0.9$,
the rotating frequency is close to $\Omega|_{\alpha=1}=-1.27$. }
\end{figure}

There is more to say about the internal modes of the vortex soliton. On one hand,
solving the Eigenvalue problem~(\ref{Eigen_vortex}) for other vorticities $m$
offers an exhaustive method for finding families of azimuthons which originate
from a vortex soliton. For example, with our method it was straightforward to
identify a rotating triple-hump azimuthon ($m=3$, $\kappa=3.8$), which rotates with
$\Omega=-\kappa/m=-1.27$. This solution bifurcate  from the $M=200$ vortex
in the Gaussian nonlocal model. The components of the generating Eigenvector are shown in Fig.~\ref{fig_azim3vor_gauss}.
Again, we can easily construct  various azimuthons with $\alpha\approx 1$  from this Eigenvector. The movie in
Fig.~\ref{fig_azim3vor_gauss} shows an example of
the propagation of the triple-hump azimuthon with $\alpha=0.9$.

It should be stressed that  not all orbitally stable Eigenvalues
can be linked to a family of azimuthons. This is obvious for Eigenvalues with
$|\kappa|>|\lambda_0|$ in the continuous part of the spectrum. Hence, we can
conclude that
$|\Omega|_{\alpha=1}|<|\lambda_0|/m$ for azimuthons bifurcating from a single-charged vortex\footnote{Please note that
the parameter $m$ determines  the number of humps of the rotating
structure.}. However, bound orbitally stable Eigenvalues do not necessary
indicate an emanating family of azimuthons. For example, the $M=200$ vortex
in the Gaussian nonlocal model possesses several other bounded internal modes
for $m=2$, e.g. $\kappa=7$. If we perturb the vortex with the corresponding Eigenvector the resulting
structure is a double-hump rotating with $\Omega=-3.5$. Closer
inspection reveals that this structure decays and hence does not belong to another family of
dipole azimuthons.
Nevertheless, since those decaying rotating structures can survive over large
propagation distances they may be indistinguishable  from true azimuthons in the
experimental conditions. This issue is currently under detailed investigation.

\section{Conclusion \label{conclusion}}

In conclusion, we analyzed properties of azimuthons, i.e.\ localized rotating nonlinear waves  in  nonlocal nonlinear
media. We showed that the frequency of angular rotation  crucially depends on the nonlocal response function. Starting
with the single charge vortex soliton and by employing  a simple variational ansatz we were able to predict accurately
the  frequencies of the rotating dipole azimuthon. In addition, we could explain how the actual shape of the response
function determines the rotation frequencies. Further on,  we computed exact dipole azimuthon solutions and their
rotation frequencies numerically and  showed that in the limits of maximal and/or minimal azimuthal amplitude
modulation, i.e., close to the dipole or vortex soliton, the rotation frequency is determined uniquely by Eigenvalues of
the bound modes of the linearized version of the respective stationary nonlocal  solution. Moreover,  the intensity
profile  of the resulting azimuthon can be constructed from the corresponding linear Eigen-solution.  This offers a
straightforward and  exhaustive method to identify rotating soliton solutions in a given nonlinear medium.

\section{Acknowledgment}
This research  was supported by the Australian Research Council. Numerical simulations were performed on the SGI Altix
3700 Bx2 cluster of the Australian Partnership for Advanced Computing (APAC).

\end{document}